\documentclass[useAMS]{mn2e}
\usepackage{psfig}

\newcommand{\targ}{IY~UMa}

\newcommand{\mnras}{MNRAS}

\newcommand{\apj}{ApJ}
\newcommand{\apjl}{ApJL}
\newcommand{\pasp}{PASP}

\title[High-speed photometry of IY UMa]{High-speed energy-resolved STJ 
photometry of the eclipsing dwarf nova \targ}

\author[D.Steeghs et al.]
{D.Steeghs$^1$, M.A.C.Perryman$^2$, A.Reynolds$^2$, J.H.J.de Bruijne$^2$, T.Marsh$^1$
\newauthor V.S.Dhillon$^3$ and A.Peacock$^2$\\
$^1$ Department of Physics and Astronomy, University of Southampton, Highfield, Southampton S017 1BJ, UK\\
$^2$ Astrophysics Missions Division, Research and Scientific Support Department, ESTEC, PO Box 299, 2200 AG Noordwijk, NL\\
 $^3$ Department of Physics and Astronomy, University of Sheffield, Sheffield, S3 7RH, UK}

\date{Accepted ;
      Received ;
      in original form}

\pagerange{\pageref{firstpage}--\pageref{lastpage}}
\pubyear{2002}

\begin{document}
\maketitle
\label{firstpage}

\begin{abstract}

We present high time-resolution photometry of  the dwarf nova IY UMa
using  the S-Cam2 super-conducting  tunnel  junction  device attached to
the 4.2m William Herschel Telescope on La Palma.
Exploiting the well-defined white dwarf  and hot spot eclipse features, we
derive an updated orbital  ephemeris for IY  UMa and an orbital period
of 0.07390897(5) days.
A white  dwarf ingress/egress duration of  $31 \pm 2$s along with the contact phases of the bright spot gives $M_1=0.79
\pm 0.04 M_{\odot}$ and $M_2=0.10 \pm  0.01 M_{\odot}$, corresponding to a
mass ratio of $q=0.125  \pm 0.008$. The  white dwarf eclipse width  of
$\Delta\phi_{\mathrm{WD}}=0.0637$ then implies $i=86.0 \pm 1^{\circ}$.
A curious rise with a duration of $30\pm 2$s is observed in the orbital lightcurves
during all  three  eclipses.  It occurs between  the end  of white
dwarf ingress and hot spot ingress  and is blue in colour.
We suggest that the source of this light lies in the buried part of
the gas stream, resulting in a compact, hot impact cavity.

\end{abstract}


\section{Introduction}

Modelling eclipses in binary systems remains one of the most
reliable tools for  extracting stellar masses and  radii.
In the semi-detached systems,   mass transfer via Roche  lobe overflow
results in an orbital lightcurve that is often dominated by light from
the accretion flow rather than the stellar components.  If viewed from
sufficiently large orbital inclinations,  the shadow cast by the  mass
donor therefore slices not  only across
the accreting stellar object, but  also probes the extended  accretion
flow around it.

\targ~ is a  member of the   cataclysmic variable (CV) subclass  of  dwarf
novae, in which a white dwarf accretes from  a late-type donor star in
an   unstable mass-transfer   rate    regime. During quiescence,   the
accretion disc fills up until a  global disc instability is triggered
and   the disc  is   emptied onto   the   primary white dwarf   during
semi-regular and short lived outbursts (e.g. Lasota 2000). 
The white dwarf has no significant magnetic field in such systems, and
the   extended accretion disc is    the  dominant contributor to  the
optical light. In  dwarf novae with  short orbital periods, longer and
more  pronounced super-outbursts  can   occur in  addition  to  normal
outbursts.  During such super-outbursts, the  tidal torque of the mass
donor star causes   the large accretion disc  to   deform and precess,
producing  prominent light  modulations (the  superhumps) at a  period
slightly longer than the orbital period of the binary.
Although IY  UMa was  already reported in  1997 to  be a variable star
(Takamizawa  1998), its status  was  only established in January  2000
after   a reported brightening.   Intensive monitoring  of this object
revealed  a  short-period  dwarf  nova  in  super-outburst  featuring
prominent superhumps as well as  deep primary eclipses with an orbital
period  of 1.77  hours  (Uemura et   al.  2000).   Since then, several
additional  outbursts    and   super-outbursts have    been  reported,
confirming  that  IY UMa  is one   of  the few  eclipsing dwarf  nova
systems.
Patterson  et al.   (2000) presented  extensive  photometry during and
after the January 2000 outburst and derived IY UMa's system parameters
using  the well-defined eclipse  features  of the  white dwarf and  hot
spot, while Rolfe   et al. (2001) used the   superhumps to trace   the
ellipsoidal shape of the accretion disc.

We obtained high time-resolution photometry of IY UMa using the S-Cam2
superconducting tunnel junction  device attached to the  4.2m William
Herschel    Telescope on La Palma (Rando et al. 2000).     This  cryogenic detector has both
intrinsic energy sensitivity and  high precision  photon time-tagging,
permitting rapid, low-resolution  spectroscopy.  It  is  thus  ideally
suited to the  study of eclipse transitions  in accreting binaries. It
consists of a  6x6 array of  tantalum superconducting tunnel  junction
(STJ) devices, cooled to 0.3 K.  The STJ  devices - hereafter pixels -
have sizes  of  25 x  25  $\mu$m$^2$, corresponding to   $\sim$  0.6 x 0.6
arcsec$^2$.   The field of  view of  the  full array  is $\sim$  4 x 4
arcsec$^2$, permitting  good background   subtraction  when the seeing   is
nominal ($\sim 1''$).  The device  has an energy  resolving power  of $\sim$ 8.5 at
500nm. Photon energies are   assigned to pulse-height   analyser (PHA)
channels in the range  0-255. The effective  bandpass - determined  by
the glass  optics, atmospheric absorption, IR filters and the  detector's quantum efficiency -
is around 340 to 680 nm, such that useful events fall between channels
60 - 165. The relationship between channel number and photon energy is
linear ($\sim$ 42.5 channels  eV$^-1$), but varies slightly from pixel
to pixel.  The arrival times of individual photons  are logged with an
accuracy of $\pm$ 5 $\mu$s using GPS timing signals.
The  data  acquisition and reduction   steps  are described in Section
\ref{reduction}, with  a detailed analysis  of the eclipses in Section
\ref{analysis} followed by our conclusions.

\section{Data acquisition and reduction}

\label{reduction}
\begin{table}
\caption[List of Observations]{Table of Observations}
\label{obs}
\begin{tabular}{ccc}
\hline
Obs. No. & UTC interval$^*$ & Phase coverage\\
\hline
1 & 21:50 - 22:30 & 0.78 - 1.15\\
2 & 22:31 - 23:11 & 1.16 - 1.53\\
3 & 23:18 - 23:38 & 1.60 - 1.79\\
4 & 23:39 - 00:19 & 1.79 - 2.17\\
5 & 01:26 - 02:07 & 2.80 - 3.18\\
\hline
\end{tabular}

\small{$^{*}$Night of April 26/27 2000}
\end{table}

We made five S-Cam2 observations  of IY UMa on the  night of Apr 26/27
2000. The detector was  mounted  on the  Nasmyth  focus of  the  4.2-m
William Herschel  telescope on La Palma, with  the  focussed light fed
through a  derotator wheel. Throughout  the night, atmospheric conditions were
good with no cloud cover and stable seeing at $\sim$ 1 arcsec, permitting good background
subtraction. Five  exposures were made on IY UMa,  covering three  eclipses (Table \ref{obs}). 

A detailed description of the data reduction procedure for S-Cam2 data
is contained  in  Perryman et al. (2001).   The first key step is  the
shifting of all  events onto  a common  energy/PHA scale, based  on  a
laboratory calibration. 
After this,  the data are  split into user-selected energy  ranges for
the  purposes  of  producing  lightcurves  in different  colour bands.
Three  energy ranges are selected,   with the boundaries between  them
selected to ensure roughly equal numbers of events in each.
For IY UMa, the adopted ranges are channels 1 - 94, 95 - 110 and 111 -
255. Using the nominal relationship  between channel number and photon
energy,  converting from eV  to nm, and then  allowing for
the  instrumental bandpass  limits, these  correspond to three  colour
bands spanning $\approx$ 325 - 470 nm, 470  - 554 nm and  554 - 643 nm
(labelled  blue,  green and red  henceforth).  In addition a composite
white light band is also produced.
In practice  -  due to  the modest  resolving power  of  the instrument -
photons  of a given   energy corresponding to  one  band have a finite
probability of spilling-over into the adjacent one.

The data are also time-binned, corrected for pixel-to-pixel efficiency
variations, corrected for  atmospheric extinction  in a time-dependent
fashion, and finally background subtracted.   The latter stage employs
two corner pixels  to obtain an estimate  of  the mean  sky background
during a  given observation, and is  implemented in a time-independent
manner.   The default  width of time   bins is  1.0s,  appropriate for
tracking eclipse  transitions in CVs, but  much shorter duration bins
may be  selected if necessary. 
Finally, the heliocentric corrections to the  individual GPS based UTC
timings  are calculated  using   the SLALIB  software  library (Wallace
2001).  This allows the calculation of heliocentric Julian dates (HJD)
for all time bins,   in order  to   compare the eclipse   timings with
previously published values.

\begin{figure*}
\psfig{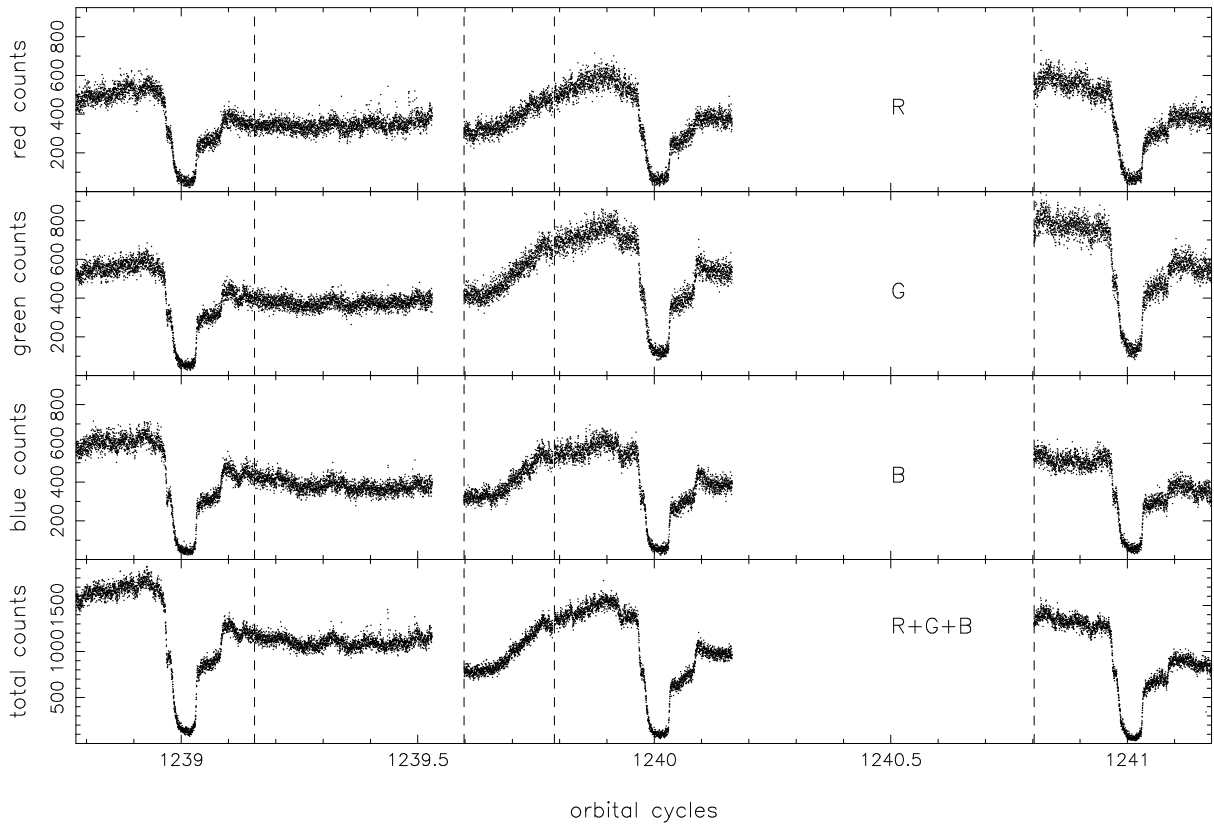}
\caption{IY UMa lightcurves in the three colour band passes referred to as {\it red} (top), {\it green} and {\it blue} as well as the combined white light lightcurve.  Plotted are object counts per second as a function of the orbital phase. Vertical lines denote the start of the individual observations.  One orbital cycle spans 106.4 minutes. }
\label{light}
\end{figure*}

\section{Data analysis}
\label{analysis}

Figure   \ref{light} displays the  extracted  object lightcurve in the
three bands as well as a white  light lightcurve with a time resolution
of 1s.  Orbital phases are calculated using the ephemeris that we will
derive later in this section. 
The  overall lightcurve  shape  is  very similar    to that  of  other
eclipsing dwarf novae such as Z Cha (e.g. Wood et al. 1986).
The out of eclipse light curve is characterised by  a broad and strong
orbital hump, caused by the impact region between the incoming gas stream and
the disc edge rotating into view.
Typical out-of-eclipse
count rates   in each  energy  band  are $\sim$350   counts s$^{-1}$,
increasing up to 600 counts s$^{-1}$ during the hot-spot hump.
The highly  structured eclipse consists of  the gradual eclipse of the
extended accretion  disc  as well as  sharp  features when the compact
white  dwarf and  hot spot  region  are eclipsed and re-appear. During
minimum light, count rates drop to $\sim$50  counts s$^{-1}$ above the
background.

The slight mismatch in counts in all three bands between the second and
third observations is due to an electronic calibration having been
performed on the array, which results in a small shift in the
overall gain of the device. This effect does not affect our
subsequent analysis.

We estimated the   visual out-of-eclipse magnitude   of IY UMa to   be
$V\sim16.8 \pm 0.5$ during our observations  based on a comparison of IY
UMa's count rate with those of observed standard stars. 
The internal variations in the lightcurve scaled as expected from
Poisson statistics, indicating that no significant systematic effects
were contributing to the variability. Uncertainties on individual data 
points were thus given by the square-root of the count rate,
delivering a typical precision of 0.05 magnitudes per data point in
the three color bands.

\subsection{Power spectrum}

\begin{figure}
\psfig{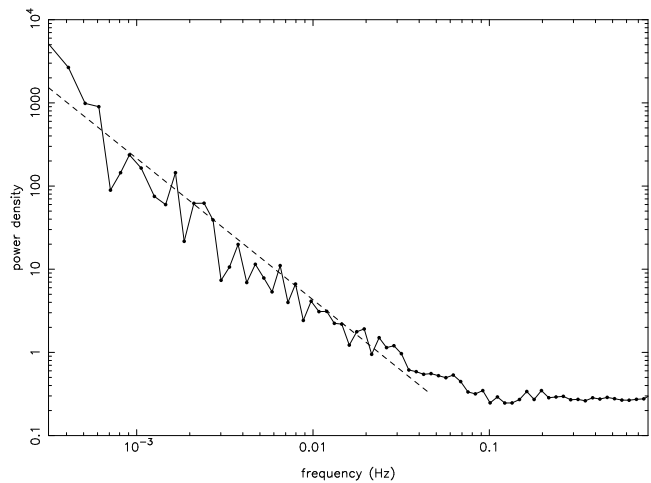}
\caption{
Power  spectrum  calculated from  the out-of-eclipse sections of the
lightcurve. All counts in  the three colour  bands were summed together
to   improve sensitivity at high frequencies.   A power law fit
resulted in a flickering index of $-1.7 \pm 0.1$ (dashed line).}
\label{power}
\end{figure}

The orbital lightcurves show, apart from  the striking eclipse and hot
spot   features,   significant aperiodic    flickering  in   all three
colours.  Flickering is   a   common characteristic of  accretion   and
reflects the dominant contribution that the accretion disc light makes
to the orbital lightcurve.
In  order to  characterise the flickering   behaviour of  IY  UMa, the
eclipse  sections ($\pm 0.1$ in terms  of phase) were removed. The data
were  then detrended with  an 8th  order  polynomial in order to remove
remaining orbital modulations and isolate the rapid flickering. 
The white light Fourier power spectrum is displayed in Figure
\ref{power}. The power spectrum has been rebinned on a logarithmic scale for plotting purposes. The  flickering  activity exhibits
itself  as a power-law component with  index $-1.7 \pm 0.1$
reaching the   Poisson   noise  level around  0.1    Hz.  

We employed different detrending filters to ascertain their effect on
the derived flickering properties. This ensured that  the power-law
index was derived from a frequency range that was not affected by the
detrending process.
No significant differences were detected  between the power spectra of
the individual color bands. We also compared the power spectrum of the
regions around the bright  spot hump (0.6-0.9) with  that of the phase
interval 0.1-0.6 and found the same flickering power-law.  Significant
flickering   power is  thus  present  on   timescales   of minutes  to
hours. 

Bruch et al. (1992) documents the flickering properties of several CVs
and  finds   flickering power-laws with  a   mean index   of  -2.  The
flickering indices span a broad range from system to system. And for a
given system  the   flickering  index  varies   significantly   between
observing epochs.  Despite  its slightly shallower flickering index, IY
UMa's flickering properties thus appear to be typical for a CV.

There is no evidence for any periodic  oscillations, such as the short
period (20-30s) dwarf nova oscillations that are ocassionally observed
in systems around  their  outburst   (e.g.   Steeghs et al  2001)   or
persistent oscillations  such as in   the  short period dwarf  nova  WZ
Sge. None have been reported for IY UMa so far.

\subsection{Eclipse timings}

Patterson et al. (2000) were able to  identify the white dwarf and hot
spot eclipses in their lightcurves towards the end of their campaign as
the system approached   its quiescent brightness.  Although  the white
dwarf  ingress  and egress was  not  well resolved, they estimated its duration to  be $\sim25$s. Our adopted time binning
of 1s is thus perfectly adequate for  resolving these features during
primary eclipse.

\begin{figure}
\psfig{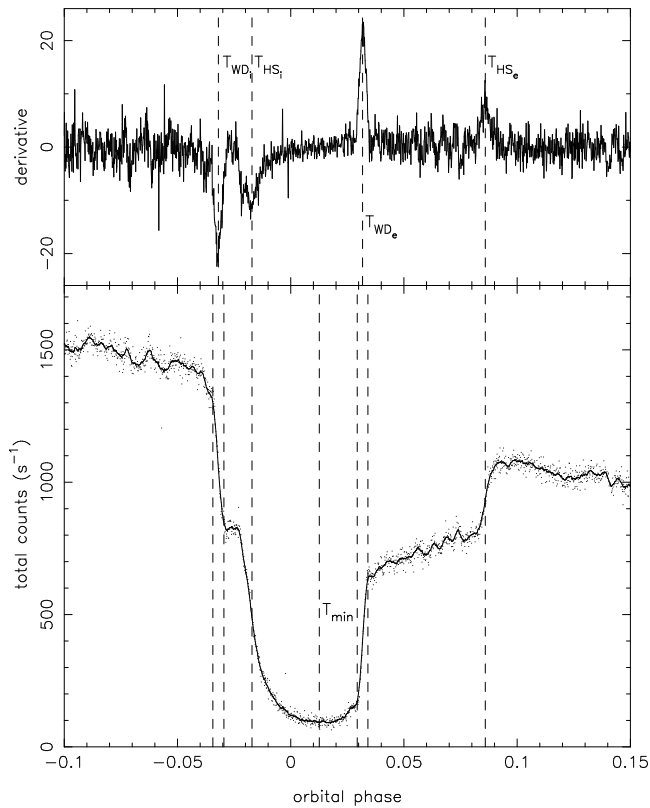}
\caption{
Close up of the IY UMa eclipse. Plotted in  the lower panel is the mean
lightcurve derived  by  adding the counts   in the three  colour  bands
together followed  by phase  binning in  order to  average the three individual
eclipses.The solid line  running through the  data points is a filtered
version of the lightcurve using a running  mean filter with width 10s.
Top  panel   is    the     numerical  derivative  of     the  filtered
lightcurce. Prominent  eclipse events are  marked with vertical lines.
}
\label{eclipse}
\end{figure}

To  illustrate the  individual eclipse  features, the  counts in the three
energy bands were added together  and the three observed eclipses were
averaged in order to  derive a mean  white light  eclipse lightcurve
for  IY UMa (Figure  \ref{eclipse}).   A smoothed lightcurve was  then
calculated using a running mean filter with a width of 10s, as well as
the numerical derivative of this smoothed lightcurve. 
The white dwarf  eclipse is well resolved  and the  start and  end of both
white    dwarf  ingress     ($T_{\mathrm{WD}{is}}$,$T_{\mathrm{WD}{ie}}$)  and      egress
($T_{\mathrm{WD}{es}}$,$T_{\mathrm{WD}{ee}}$) were  measured  for all three eclipses. 
White dwarf ingress/egress ($T_{\mathrm{WD}_i}$,$T_{\mathrm{WD}_e}$) was then calculated
as the mid-point between the beginning and end of ingress/egress. As a
complimentary measurement of  white dwarf ingress/egress, the point at
which  the derivative is maximal ($T_{i_{deriv}}$,$T_{e_{deriv}}$) was
also determined.   The derived white  dwarf  ingress/egress times when
comparing  the two methods always agreed  to  better than 2s, which we
use as a conservative error  estimate on all individual eclipse timing
measurements.

Immediately after white dwarf ingress and before the start of the bright
spot ingress,  an increase in the count  rate is observed for  a short
duration.  Since this curious eclipse feature  may influence the start
of hot spot ingress,  hot spot ingress  and egress times were measured
instead as     the phases of   maximum  derivative   ($T_{\mathrm{HS}_{i}}$ and
$T_{\mathrm{HS}_{e}}$).  Finally,  the phase of minimum  light was measured for
each    eclipse ($T_{min}$).  We   could  not  measure any significant
differences between the eclipse timings in the three energy bands, and
therefore  Table \ref{timings}  lists   all measured timings based  on
summed  white-light  lightcurves  in order  to  improve  the signal to
noise.

\begin{table}
\caption[Eclipse timings]{Eclipse measurements}
\label{timings}
\begin{tabular}{p{2.5cm}ccc}
\hline
eclipse number:& 1239 & 1240 & 1241 \\
\hline
$T_{\mathrm{WD}_{is}}$& 2.4244215 & 2.4983506	& 2.5722487	\\
$T_{\mathrm{WD}_{ie}}$& 2.4247713 & 2.4987550	& 2.5725796	\\
$T_{\mathrm{WD}_{i}}$& 2.4245964 & 2.4985528	& 2.5724142	\\
$T_{\mathrm{WD}_{es}}$& 2.4291654 & 2.5030394	& 2.5769555	\\
$T_{\mathrm{WD}_{ee}}$& 2.4294779 & 2.5034435	& 2.5772865	\\
$T_{\mathrm{WD}_{e}}$& 2.4293216 & 2.5032414	& 2.5771210	\\
$T_{i_{deriv}}$& 2.4246092 & 2.4985483	& 2.5724325	\\
$T_{e_{deriv}}$& 2.4293492 & 2.5032415	& 2.5771346	\\
$T_{\mathrm{HS}_{i}}$& 2.4257624 & 2.4995868	& 2.5735045	\\
$T_{\mathrm{HS}_{e}}$& 2.4332554 & 2.5072813	& 2.5811620	\\
$T_{min}$& 2.4278598 & 2.5018623	& 2.5757053	\\
\hline
\end{tabular}

\small{all times are HJD - 2451660.0, uncertainty $\pm 2 \times 10^{-5}$}
\end{table}

Our      three       accurate             mid-eclipse            times
($T_0=1/2(T_{\mathrm{WD}_{i}}+T_{\mathrm{WD}_{e}})$) together  with  those listed in   Patterson et al. (2000) were used  to calculate a refined orbital
ephemeris for IY UMa;
\[
\mathrm{T_{mid-eclipse}} = \mathrm{HJD~} 2451570.85376(2) + 0.07390897(5) \mathrm{~E}
\]
\noindent 
All  eclipse measurements  were given equal  weight  in this ephemeris
determination. The  figures in brackets denote the   error on the last
digits. The above ephemeris is used  throughout this paper whenever an
orbital phase needs to be calculated.
In    order  to   ascertain the   reliability    of our   white  dwarf
ingress/egress measurements   as  given  in  Table \ref{timings},   we
calculated synthetic white    dwarf lightcurves using a spherical
white dwarf with given linear limb darkening coefficient.
The limb of the Roche lobe was evaluated using a three-point
approximation (e.g. Wood \& Horne 1990). White dwarf eclipse lightcurves were then calculated using
variable limb darkening coefficients and ingress/egress phases were
measured in the same manner as the data.
We  thus
established that   the individual phase  measurements  are reliable to
within 4 seconds, with no measurable systematic offset.
We then derived our final  mean white dwarf ingress/egress duration of
$31 \pm 2$ s based  on 6 individual measurements. The
total  white dwarf eclipse  duration  is $407 \pm  1$  s, or 0.0637 in
terms of phase.
Minimum light occurs $80 \pm  3$ seconds after white dwarf conjunction
reflecting the distortion  caused by the  hot spot, while  the curious
rise in the eclipse after  the end of  white dwarf ingress reaches its
peak $30 \pm  2$  seconds later. 
The duration of white dwarf ingress and egress are indistinguishable,
indicating that the white dwarf is fully eclipsed before the rise of
the short duration feature commences.
We   will look into  this eclipse feature in more detail in Section 
\ref{blibsec}.

\subsection{System parameters}

We follow the techniques of  Wood et al.   (1986) and Patterson et al.
(2000),  and use the   measured hot spot  ingress/egress  phases  as a
constraint on  the binary mass  ratio $q=M_2/M_1$.  Predicted hot spot
ingress and egress phases were calculated as  a function of mass ratio
using ballistic stream trajectories in  the binary potential, assuming
the secondary  star  is filling its Roche  lobe.  The distance between
white  dwarf  and   the  first Lagrangian  point  is  referred  to  as
$R_{L_1}$.  Using our three hot spot phases, we find that $q=0.125 \pm
0.008$, while  $R_{disk}=0.54  \pm 0.01 R_{L_1}$  which corresponds to
$\sim 0.38$ in terms of the binary separation $a$.
Our derived   mass ratio agrees  with, and improves upon,   the value
derived by Patterson et al.  (2000). IY UMa  is one of the calibrators
for Patterson's empirical relation between the superhump period excess
($\epsilon$) and the  binary mass ratio  (Patterson  2001).  We remark
that our value for  the mass ratio  together with the observed  period
difference    between  orbital      period   and  superhump     period
$\epsilon=(P_{sh}-P_{orb})/P_{orb}=0.0281$ leads  to   $\epsilon/q=0.22
\pm 0.01$ based on IY UMa alone if we assume a linear relation.

Interestingly,   our  inferred accretion  disc  size  is significantly
larger  compared to the   $R_{disk}=0.28a$  measured by Patterson   et
al.  (2000)  and     $R_{disk}=0.25a$ derived by    Stanichev   et al.
(2001). Their observations took place 1-2 months  before our data, and
thus closer to the superoutburst that was first reported on January 13, 2000.
We estimated the brightness of IY UMa to be $V \sim 16.80$ during
our observations.  This   is brighter than  the  reported  quiescent V
magnitude of $\sim 17.5$, which suggest that another outburst may have
occured around the  time of our  observations.  On the other hand, the
appearance  of the  light   curve  is very  much  like   the quiescent
lightcurves reported  by  Patterson et al.   which  suggests that  our
observations would  need to be  during the  later  phases  of such  an
outburst.  Given  the faintess of IY  UMa, normal outbursts are easily
missed by  the variable star networks and  it it thus no surprise that
not  all brightenings are reported.   This is precisely the reason why
IY UMa,    which reaches $V\sim14$  during   super-outburst,  was only
recently discovered.
The  occurence of a normal outburst,  rather  than a super-outburst, in
between our time of observations and that of Patterson et al.  (2000) 
could explain the observed  disc size.  At  the start of an  outburst,
the disc expands   rapidly and then   gradually  shrinks back  to  its
quiescent  value. A large  disc is thus  a tell  tale sign of enhanced
mass accretion through the disc due to an outburst.

Having established the duration of white  dwarf eclipse, we can then
exploit the unique  relation    between $q$, the  binary
inclination    $i$     and  the   white      dwarf   eclipse  duration
$\Delta\phi_{\mathrm{WD}}=0.0637$ and derive $i=86.0 \pm 1^{\circ}$.
Finally, if we    assume that the   white dwarf  radius is given    by
Eggleton's  mass-radius relationship  (Verbunt  \& Rappaport 1988), we
can use the measured   white dwarf ingress/egress duration  and determine
the masses of  the  stellar  components. 
We then find for the masses of the white dwarf  and mass donor star in
IY   UMa;  $M_1=0.79 \pm  0.04     M_{\odot}$ and $M_2=0.10  \pm  0.01
M_{\odot}$.  These  values are very  similar  to those  of other short
period  systems such as Z Cha  and HT Cas, and as   such typical for a
short period CV with a mid to late M-type secondary star (Smith
\& Dhillon 1998 ; Rolfe, Abbott \& Haswell 2002).

\subsection{The rising eclipse feature}
\label{blibsec}

\begin{figure}
\psfig{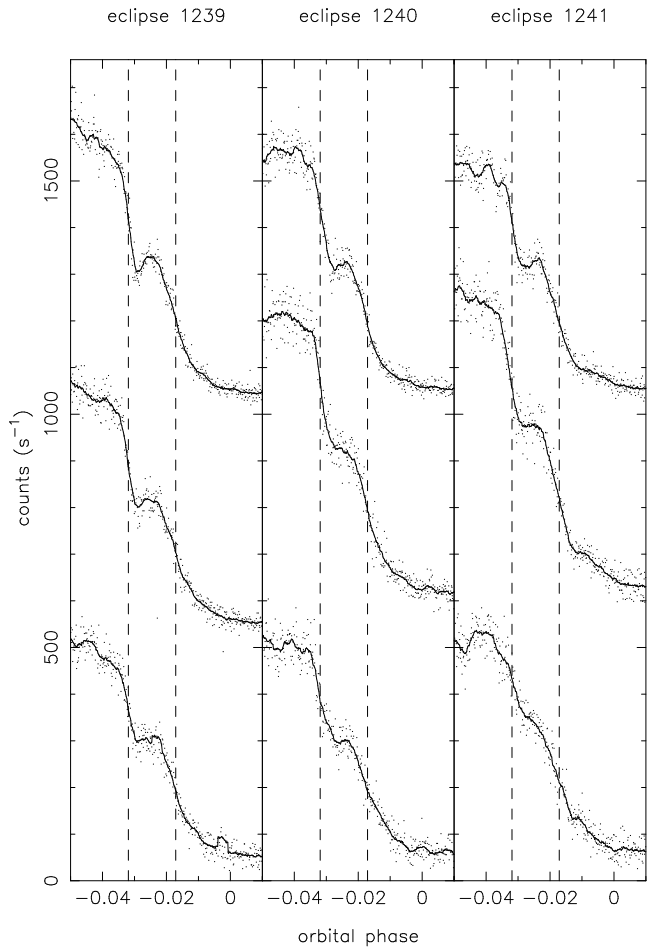}
\caption{ 
A  closer  look  at the  interval   between white dwarf   and hot spot
ingress. Plotted  are the three  individual eclipses in the three colour
bands. The green and  blue (top) lightcurves are  offset by  500 and 1000 counts, respectively, for plotting purposes. Vertical lines  denote mid white dwarf and hot
spot   ingress phases. Running  through  the data points  is a filtered
version of    the lightcurve using  a   running mean filter   of width
10s.    }
\label{blib}
\end{figure}

\begin{figure}
\psfig{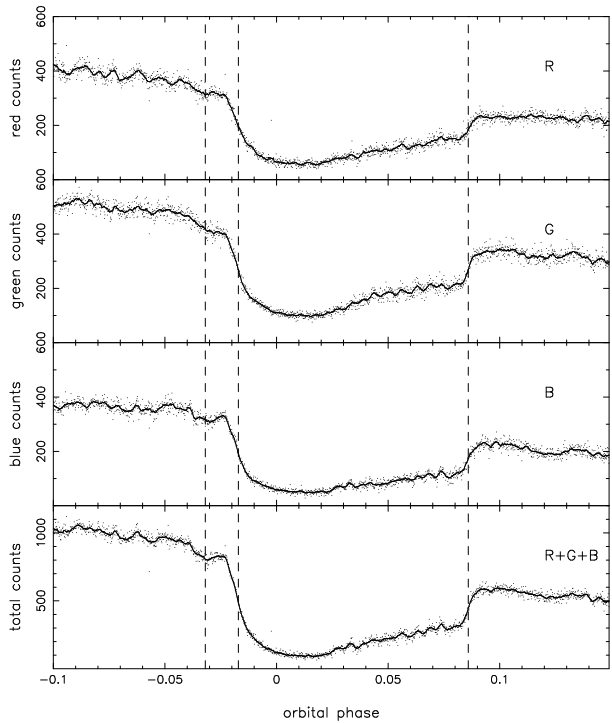}
\caption{The mean eclipses of IY UMa after subtraction of the white dwarf contribution. Marked are white dwarf ingress and hot spot ingress/egress phases. Solid line is a running mean filtered version of the time series.}
\label{wdsub}
\end{figure}

We remarked earlier that the end of white dwarf ingress is followed by
a short ($\sim 30$s) rise in  the lightcurve before the hot spot
is eclipsed.  This is surprising since more  and more of  the accretion
flow is  falling  into the shadow of  the  secondary star which should
result in a monotonically decreasing eclipse lightcurve.
In  order  to understand the  nature   of this lightcurve  feature, we
illustrate  its properties in more  detail  in Figure \ref{blib}.  The
feature is visible in all  three observed eclipses  and is always most
pronounced in the blue band, and weakest in  the red lightcurves. This
suggests that the source responsible  for this  feature is compact  and
hot, with  similar colours to the hot  spot. Its timing lies
exactly   between   the end of   white dwarf  ingress ($T_{\mathrm{WD}_{ie}}$) and hot   spot ingress ($T_{\mathrm{HS}_{i}}$).   However, no similar feature  is visible between white dwarf
and hot spot egress, implying that the source of this additional light
is not visible when observed from the opposite direction.
This suggests that some feature rotates into view just after the white
dwarf is eclipsed,  while it  is  not  visible at most  other  orbital
phases. 
In order to facilitate the interpretation, we modeled the contribution
of the white dwarf for  each color band  and subtracted its light from
the orbital lightcurve. To WD flux was assumed to be constant before
ingress and after egress and zero during white dwarf eclipse. The
lightcurve was then completed by integrating the numerical
derivative of the filtered lightcurve during white dwarf ingress and
egress (ala Wood et al. 1986). 
  In Figure \ref{wdsub}, we  plot the mean white
dwarf  subtracted  lightcurves  for all   colours.  The rising feature
reveals  itself again most notably   in the blue   band, with an  even
higher blue-to-green ratio than the bright spot.
Its surface brightness  needs to be larger than  that of the hot inner
disc regions since the count rate increases while the inner disc falls
in the shadow of the secondary star. 

\begin{figure}
\psfig{figure=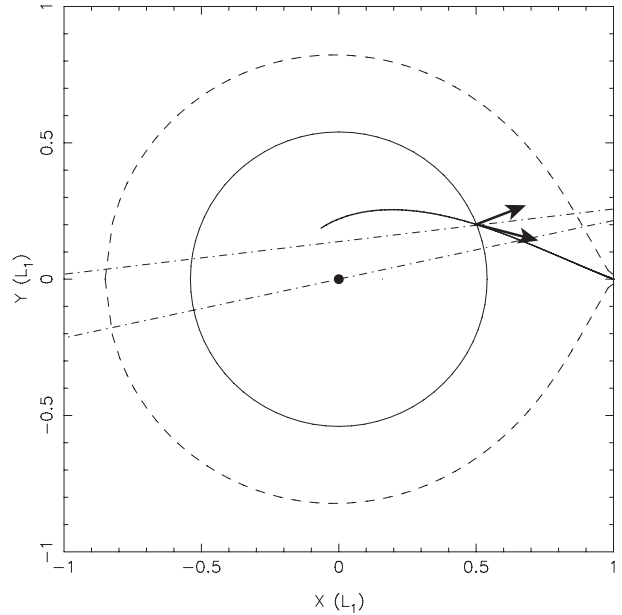,width=8cm}
\caption{Pictured is the accretion stream in the primary Roche lobe for IY UMa's system parameters. Solid circle represents the deduced disc radius at the time of our observations. Dot-dashed lines denote the edge of the secondary shadow at white dwarf  and hot spot ingress. Arrows indicate the direction of emission of material beamed perpendicular to the hot spot impact point at the disc edge compared to beaming along the stream direction.   }
\label{geometry}
\end{figure}

To illustrate the
geometry in IY UMa near these orbital phases, we plot in Figure
\ref{geometry} 
the Roche geometry for  the system parameters of  IY UMa. The inferred
disc size is indicated by the solid circle, and the canonical location
of  the  hot spot  is the  intersection  of  the  ballistic gas stream
trajectory with the disc edge. The path of the secondary star's shadow
is also indicated as  it advances between white  dwarf ingress and hot
spot ingress.

A likely possibility  is that the gas stream  on impact buries itself
(partly) into the disc, creating a narrow, hot impact cavity.  If emission is
beamed  preferentially along the  direction of  the  stream near
impact  rather than perpendicular to  the disc edge, this buried part
of the stream would then make a growing contribution to the lightcurve
after white dwarf  egress as the angle  between the stream  trajectory
and    the   line   of sight     continues  to  decrease    (arrows in
Fig. \ref{geometry}).
The resulting rise in  the lightcurve is  cut short when the shadow of
the secondary covers the stream impact region and the hot spot ingress
takes place, explaining the short  duration of the feature.  This  hot
cavity along  the stream trajectory is also  consistent  with its blue
colour.  Since  the cavity  is  buried in the  disc,  it would only be
visible  when the  line of  sight looks into  the  cavity, and not  at
opposite  orbital phases. This qualitative  model can thus explain all
the observed features.

No similar  feature has been observed  in other eclipsing CVs, but the
very high binary inclination of IY UMa  may be a prerequisite for such
line of sight effects to   occur.  The observable properties of   this
stream  cavity effect  depend   mainly  on  how tightly and   in what
direction the light is beamed.   The  detailed geometry of the  stream
impact region is notoriously complex with possible other complications
such as stream overflow and  a vertically and azimuthally extended hot
spot down-stream  along  the  disc edge.  Our  data suggest  that the
location of  highest dissipation  may not be   at the disc  edge,  the
classical location of the bright spot, but further into the disc.
With more good quality high-speed observations  of eclipsing CVs, such
models can be more thoroughly tested.

\section{Conclusions}

We presented high-time resolution photometry  of IY UMa covering three
of its deep eclipses.  The data were  obtained $\sim3.5$ months after the
January 2000 super-outburst using the S-Cam2 STJ-based detector array.
Using the well defined white-dwarf eclipse features, and our high time
resolution, we derived an updated orbital ephemeris for this eclipsing
dwarf nova system.  In conjunction with measurements of the hot spot eclipse
phases, we were able to derive accurate system parameters for IY UMa.
We found $M_1=0.79 \pm  0.04 M_{\odot}$ and $M_2=0.10 \pm 0.01 M_{\odot}$, 
corresponding to a mass ratio of $q=0.125  \pm 0.008$. The white dwarf
eclipse width  of $\Delta\phi_{\mathrm{WD}}=0.0637$  then implies   $i=86.0 \pm
1^{\circ}$.

The measured  disc radius (   $R_{disk}=0.54  \pm 0.01  R_{L_1}$  ) was
considerably larger than the values reported towards the end of the January
2000 super-outburst. This is most  likely  due to  the occurence of  a
normal outburst just before our observations. The observed count-rates
suggest that  IY UMa was slightly  brighter than its nominal quiescent
magnitude, supporting the occurence of a mass accretion rate increase.

A curious  short-lived  rise in the  orbital  lightcurve  was observed
during all  three  eclipses.  It occurs between  the end  of white
dwarf ingress and hot spot ingress  and has a blue colour.
We suggested that the source of this light lies in the buried part of
the gas stream, resulting in a compact, hot impact cavity.
The curious  line-of-sight effects leading  to this lightcurve feature
are  the  result of   the very   high binary  inclination  of IY  UMa,
making it a wonderful system for exploring the complex geometry
of the accretion flow in dwarf novae.

Given sufficient   time resolution  and signal  to   noise, lightcurve
analysis of    high inclination systems  such as   \targ~ are thus an
invaluable tool.  The future looks  bright with the energy resolution of
STJ devices significantly improving,  and large-aperture telescopes providing sufficient photons to perform high-time resolution spectroscopy.

\section*{Acknowledgments}

DS is supported by a PPARC Fellowship.  The William Herschel telescope
is operated on the island of La Palma by the Isaac Newton Group in the
Spanish Observatorio del Roque de  los  Muchachos of the Instituto  de
Astrofisica de Canarias.

\label{lastpage}
\end{document}